# ICT USE AND LIVELIHOODS OF WOMEN MICROENTERPRISES IN MALAWI

Donald Flywell Malanga, University of Livingstonia, dmalanga@unilia.ac.mw

Memory Banda, University of Livingstonia, mbanda@unilia.ac.mw

**Abstract:** This study aimed to assess the impacts of ICTs on livelihoods of women microenterprises in Malawi. The study was an interpretive qualitative approach in which semi-structured interviews, observation and field notes were used to collect data. About 25 women involved in various microenterprises in three rural areas of Karonga district in Malawi were purposively selected to participate in the study. The framework for the study was based on Serrat's sustainable livelihood approach. The study noted that the use of ICTs potentially enabled women microenterprises to build their financial, human, social, and informational capital assets. The study found that ICTs to some extent contributed to the livelihoods of women microenterprises such as improved access to information; diversification of business opportunities, improved communication, improved marketing, and reduced transport costs. As a result, this led to sustainable use of resources, improved well-being, and empowerment for women. However, unreliable electricity; lack of affordable ICT devices; lack of awareness to utilise ICTs in businesses and lack of ICT literacy skills were major concerns that affected women microenterprises. The study offers insights to research practioners, policy makers and other stakeholders on the role of ICTs in fostering women microentrepreneurship in rural communities of Malawi.

## 1. INTRODUCTION

Empowering women through microenterprises in developing countries is considered a vital issue for socio-economic development (Osman, Malanga & Chigona, 2019b; Makoza & Chigona, 2012). A recent global study on entrepreneurship found that women are the majority among the upcoming microenterprise owners in the developing world (Azerbaijan, 2016). For instance, in Sub-Saharan Africa, the female-to-male microenterprise ratio is very high in comparison with North Africa and the Middle East (Kelly, et al., 2016). Women are far more likely to be in self-employment as opposed to being employers or wage workers. Further, women entrepreneurs are more likely than men to be in services and in traditional lower value added sectors such as food processing, and garments (Qureshi & Xiong, 2017). Women are also less likely to register their business than men (Hallward-Driemeier, 2011). This gender disparities in microenterprises underscore for more investment and diversified women microentrepreneurship (Osman, Chigona & Malanga, 2019a). This is because such microenterprises arguably create incomes, jobs, skill development, alleviate poverty, and eventually crucial agents of change in the informal economy (Afrah & Fabiha, 2017).

Microenterprises are simply described as businesses owned by less than five employees (Kelly, et al., 2016; Abor & Quartey, 2010). They are characterised as being informal, survivalist, low revenue, no business records, and have no clear separation of business and personal finances (Makoza &

Chigona, 2011). However, studies also indicate that the majority of women engaged in such microentrepreneurship face a number of setbacks such as limited resources, capacity and access support (Adams, 2012).

The situation is further excercabated by cultural norms and values, high women reproductive workload and children concerns, and low education levels (Lwoga & Chigona, 2020). Further, the repressive laws and customs rooted in cultural beliefs make women's ability to manage property and conduct business hindered, and sometimes they cannot travel without their husbands' consent, thus making women microenterprise chances of survival limited (Qureshi & Xiong, 2017).

Prior studies suggest that as Information and Communication Technologies (ICTs) continue to diffuse, and greater attempts are made to apply them to microenterprises to enhance rural-poverty reduction agenda (Makoza & Chigona, 2012; Duncombe & Heeks, 2005). In this study, ICTs are defined as electronic means of capturing, processing, storing and communicating information. These may include modern digital ICTs (such as internet, computers, social media, and mobile phones, etc.) and tradition ICTs such as radios, televisions (TVs), and landline telephones (Duncombe & Heeks, 2005).

In comparison to the developed world, few microenterprises in developing countries have direct access to modern digital ICTs. The majority of women microenterprises possess traditional ICTs such as radios, with limited access to personal landline phones and television due to high cost (Qureshi & Xiong, 2017; Duncombe, 2006). However, since the majority of women microenterprises operate in rural areas, application of ICTs into their businesses could potentially address some of the problems (Osman, et al., 2019b; Sife et al., 2017). In this regard, ICTs such as mobile phones, social media, internet, traditional radios and TVs etc can potentially support the livelihoods of women microenterprises through increased labour productivity, increased income, better communication, and better access to information and reduced costs (Osman et al., 2019b; Good & Qureshi, 2009).

In Africa, studies have found that use of ICTs in microenterprises assist in reducing information failures that impact on investment decisions and business activities (Esselaar, Stock, Ndawalana &Deen-Swarry, 2007; Duncombe &Heeks, 2005). Besides, studies have shown that women that adopt ICTs such as mobile phones in their business activities, more often, are able to keep in contact with customers and clients compared to any other form of communication (Esselaar et al., 2007). On the other hand, full optimal use of ICTs in women microenterprises is beset by myriad challenges. These include low education, lack of access to financial credit, lack of access to markets, lack of ICT skills for use in business, repressive cultural norms, lack of access to affordable ICTs such as smartphones (Qureshi & Xiong, 2017; Charman, 2016).

In Malawi, anecdotal reports suggest that the number of women using ICTs in their business activities is burgeoning. However, there is paucity of empirical studies on the impact of ICTs on the livelihoods of women microenterprises in the country. The majority of studies available have focused much on adoption of ICTs by small and medium enterprises (SMEs), and has paid little attention to microenterprises (Makoza & Chigona, 2012). Due to uniqueness of women microenterprises, the findings reported from such SMEs studies cannot be generalized to microenterprises, particularly for those owned by women. Moreover, studies focusing on adoption of ICTs by SMEs alone have been critised for being limited in scope because they do not take into account human development aspects such as livelihoods (Osman et al., 2019a; Makoza, 2011), which was the main objective of this study. To bridge this knowledge, the women main research question posed for the study was: How do ICTs impact on the livelihoods of rural women microenterprises in Malawi? The study aimed to answer the following four sub-questions:
- How do ICTs help women microenterprises deal with vulnerabilities?

- How do women microenterprises use ICTs to strengthen utilisation of their capital assets?
- How do ICTs strengthen the livelihood structures of women microenterprises?
- What are the effects of ICTs on livelihood outcomes of women microenterprises?

To address these questions, the study was guided by sustainable livelihood approach (SLA) (Serrat, 2017) as a theoretical lens. The paper is arranged according to the following sections: background to the study, analytical framework, methodology, results and discussion, and conclusion.

## 2. BACKGROUND TO THE STUDY

Malawi gained its independence from Great Britain in 1964. It borders Tanzania, Zambia and Mozambique. The country has an estimated population of 17.7 million people of which 85% live in rural areas (National Statistics Office, 2015). It is classified as one of the least developed countries in the world with a Gross Domestic Capital per Capita is USD 516.80 (FinMark Trust & Genesis Analytics, 2019). Most women are working in agricultural sector which is a backbone of Malawi's economy. Of those in non-agricultural waged employment, 21% are women and 79% are men and the numbers have remained the same over the years (National Statistics Office, 2015). The root causes points to culture, unequal power relations between men and women, which ensure male dominance over women. The unequal status of women is further exacerbated by poverty and discriminatory treatment in the family and public life (Spotlight Initiative, 2020).The country faces many challenges such as food security, high rates of unemployment, impact of HIV/AIDs, high illiteracy levels, and extreme poverty (National Statistics Office, 2015). The overall mobile penetration is estimated at 45.5%, while internet is 6.5% below the recommended threshold of 19% by International Telecommunication Union (Malanga & Chigona, 2018; Malanga, 2017). About 34.5% of women own a mobile phone, 0.6% own a desktop computer, 1.8% own a Laptop, while just 4.7 % of them have access to the internet (Malanga & Chigona, 2018). The low rate of ICT penetration in Malawi is attributed to the country's weak economy, high value added tax (VAT) imposed on importation of ICT gadgets and other contextual factors (Malanga & Kamanga, 2019).

### 2.1 State of microenterprises in Malawi

The Malawi government has long recognised the role of SMEs play in socio-economic development. This was evidenced by passing a Small and Medium Enterprises (SMEs) policy in 1998, which was later revised in 2018 to include microenterprises (FinMark Trust & Genesis Analytics, 2019). In Malawi context, microenterprise is defined as business with less than four employees (Darroll, 2012). Literature indicates that the number of people especially women who cannot find formal employment in the country, find themselves in microenterprises as alternative means to support their livelihoods (Makoza & Chigona, 2012).

Recent study report by FinScope (2019) estimates that about 59% of microenterprises are individual entrepreneurs who are not employers. While the remaining 41% generate employment. Further, 87% of microenterprises are retailers while the remaining 13% render professional and skilled services. Based on location, 85% of microenterprises are located in rural areas and 15 % are located in urban areas. The study further reports that in Malawi, 54% of microenterprises are owned by males while 46% are owned by females. Consistent with previous studies, more males in Malawi are likely to run large businesses, while women are more likely to be individual entrepreneurs (Porter et al., 2020; Makoza & Chigona, 2012).

Despite the benefits associated with microenterprises in the country, their survival is curtailed by a number of setbacks. The challenges stem from limited access to finance, limited access to information, limited access to ICTs, limited access to technologies, inadequate infrastructure and utilities, and lack of business networking (FinMark Trust & Genesis Analytics, 2019; Makoza & Chigona, 2011).

## 2.2 Formal support institutions for microenterprise in Malawi

Despite the challenges facing SMEs in the country including women microenterprises, Malawi government has been spearheading the establishment of various formal support institutions (FinScope, 2019). Some of the notable ones include Pride Africa, Small and Medium Enterprise Development Institute (SMEDI), Youth Enterprise Development Fund (YEDF), Malawi Confederation of Chambers of Commerce and Industry (MCCCI) (FinScope, 2019). Examples of support programmes and services include business skills training, mentoring, access to credits, loans, business management, etc (Malanga &Kamanga, 2019; Makoza & Chigona, 2011).

## 3. ANALYTICAL FRAMEWORK

The study was guided by sustainable livelihood approach (SLA) postulated by Department for International Development (DFID) in 2000, and operationalised further by Serrat (2017). The SLA is widely used in the field of development, and increasingly in the context of ICTs based development initiatives (Chilimo & Ngulube, 2011). The use of SLA in this study was useful because bridging the rural-urban digital inequality is not merely about increasing the number of ICTs access and affordability, but it is also about impacting the lives of people such as women microenterprises, and empowering them through ICTs (Serrat, 2017; Sife at al., 2017).

### 3.1 Concepts of sustainable livelihood approach

The SLA consists of a number of elements which are used to holistically analyse the link between issues and activities within the livelihood. These include vulnerabilities, assets, structures (social relations, organisation and institutions), strategies and outcomes. Figure 1 illustrates the interactions between the various elements of SLA.

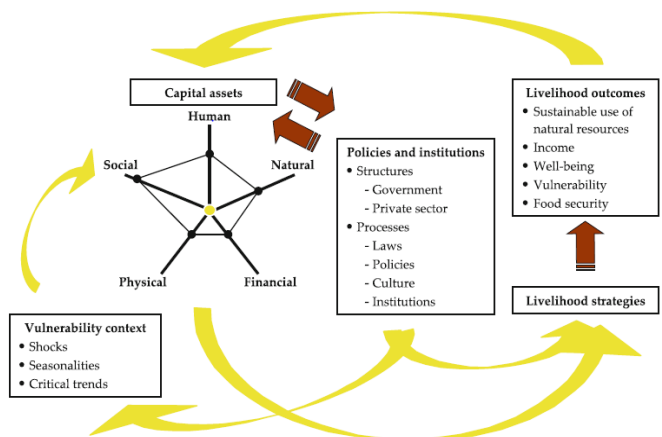

**Figure 1: sustainable livelihood framework (Serrat, 2017)**

### 3.1. 1. Vulnerability context

Vulnerability is characterised as insecurity in the well-being of individuals, households, and communities in the face of changes in their external environment (DFID, 2000). In this regard, vulnerabilities are simply defined as external factors affecting people's livelihoods, which lead to hardship (Duncombe, 2006). Vulnerabilities take three forms, namely stress, shocks and seasonality. Stress are long term trends that affect people, such as conflicts, declining natural resources, climate change and social exclusion. Shocks are conditions or events that are sudden and unpredictable, such as epidemics and natural disasters. Seasonality refers to changes in price of commodities and shifting of employment opportunities (Serrat, 2017).

### 3.1. 2. Capital assets

Capital assets are resources that households have access to and use them to produce goods or services as a means of sustaining their lives. Increased access to assets may lead to more sustainable livelihoods (Chilimo & Ngulube, 2011). There are various forms of assets and they include: (i) Human capital: the knowledge and skills that people have and use to achieve sustainable livelihoods. (ii)Social capital: these include social relations, membership to organisations. (iii) Natural capital: they include land, water, wildlife and biodiversity. (iv) Financial capital: Financial capitals are resources that can be used to establish livelihood activities such as savings, cash and access to loans. (v)Physical capital: resources created through the economic production process e.g. roads, power lines and supplies (Sife et al., 2017).

### 3.1.3. Transforming structures and processes

Structures are the public and private organisations and institutions that facilitate the attainment and use of capital assets through implementation of policy and legislation, delivery of services that affect the livelihoods (Serrat, 2017). Examples of organisations and institutions are Government Departments, Non-governmental Organisations (NGOs) and Community Based Organizations (CBOs) that deliver services for livelihood to communities including microenterprises owned by women. Processes embrace the laws, regulations, policies, operational arrangements, agreements and societal norms that in turn determine the way in which the structures operate (Serrat, 2017; Chilimo &Ngulube, 2011).

### 3.1.4. Livelihood strategies and outcomes

Strategies are activities that generate a means to achieve sustainable livelihood. Strategies can be implemented by the household in the form of economic activities, or by the institution coming up with interventions that affects the livelihood of households. These strategies may change all the time to respond to the factors affecting livelihoods (Qureshi & Xiong, 2017).

Livelihood outcomes are the results of applying livelihood strategies and use of capital assets. They include increased wellbeing, reduced vulnerability, improved food security, recovered human dignity and more sustainable use of resources (Chilimo &Ngulube, 2011). Besides, livelihood outcomes may further lead long term changes that involve determining the extent to which ICTs can support to mitigate the vulnerabilities and expand or diversify the existing activities of microenterprises (Serrat, 2017).

## 4.　　METHODOLOGY

### 4.1 Research approach

This study adopted a qualitative interpretive philosophical approach. Interpretivism assumes that human beings as social beings are different from physical phenomena because they create meanings that stem from multiple realities (Sanders, Lewis & Thornhill, 2014). Because of the context nature of the study, interpretive approach was adopted to help create new and richer understanding on the meanings that women attached to use of ICTs in their business activities. We employed a multi-case study design (Creswell, 2014). The use of multi-case study helped the researchers to explore in depth and understand the behavioural conditions that women operated their microenterprises on those selected case study areas (Creswell, 2014).

About 25 microenterprises owned by women were targeted and sampled purposively at Uliwa, Nyugwe, and Mlare rural areas of Karonga district in Malawi. Women were targeted from these rural communities because of their high involvement in various business activities aimed at improving their livelihoods and well-being. Semi-structured interviews, filed notes, and observation were used to gather data for the study. These data collection instruments accorded the researchers the opportunity for immediate response from participants and the opportunity to ask immediate follow-up questions (Kumar, 2011). Questions asked to research participants were based on existing literature (Mbuyisa, 2017; Makoza, 2011; Chilimo, 2008), and were adapted to suite the context of the present study.

### 4.2 Site Selection: Karonga district

This study was conducted in three selected rural areas situated south of Karonga district: Uliwa, Nyungwe, and Mlare under Traditional Authorities Mwirang'ombe and Wasambo respectively. All these selected areas are situated along the Mzuzu city to Karonga M1 road. The tar marked road provides means of transports and connect women microentrepreneurs from rural markets to markets main townships of Karonga district and the Mzuzu city respectively (MDHS, 2018).

Karonga district is situated 220km North of Mzuzu city and 50km south of Songwe border with Tanzania (Malanga & Kamanga, 2019). It has a population of 365,028 representing 1.3% of the country's population. Women accounts for 51.7% of the population. Number of households is estimated at 74,953 in the district (MDHS, 2018). The majority of women are involved in subsistence agricultural activities as their main source of household livelihoods. Household ownership includes TVs, radios, landline, satellite dishes, mobile phones, computers and refrigerators, etc (Malawi Demographic and Health Survey [MDHS], 2018).

### 4.3 Case selection criteria

In terms of selecting the cases, we adapted the criteria postulated by Qureshi and Xiong (2017). The authors suggest a number of criteria that should be taken into consideration when selecting microenterprises as unit of analysis. As indicated in Table 1, we employed similar variables to select research participants for this study.

| Criterion | Description |
|---|---|
| Existing challenge | The microenterprise is facing challenges in operation such as lack knowledge, lack of resources, and lack of skills |
| Target | Microenterprise should be owned by a woman/female |
| Potentiality | The microenterprise should:<br>a.     have a business potential to grow and expand her business by usage of ICTs<br>b.     have enough funds to invest in ICTs<br>c.     desire to gain access to new markets |
| Ownership | The ownership of a microenterprise should be sole proprietorship or partnership |
| Scale of the business | The microenterprise should follow the criteria of the general definition of employing less than five employees based on developing countries context |
| Years of operation | The microenterprise should exist for more than one year, so that endogenous variables lead to the challenge, e.g., lack of cash flow. |
| Geographical location | The microenterprise should be situated in a local or rural setting |

**Table 1: Case selection criteria adapted from Qureshi and Xiong (2017)**

### 4.4 Research ethics

Permission to conduct the study was obtained from the individual research participants. Research participants were informed that their involvement in the study was voluntary. To ensure that participants' privacy and confidentiality are protected, all recordings and transcripts were stored securely. Pseudonyms were used to protect their identities. In addition, oral consent was also obtained from their village leaders.

### 4.4 Data collection and analysis

Thematic analysis was used to analyse the data collected from this study. We used a codebook and smartphone to record the interviews. The codebook was developed based on SLA that guided this study. Thematic analysis was used due to its ability to condense important attributes of large amounts of data, recognize, and analyse patterns of significance and meaning in a data set (Braun &

Clarke, 2006). Thematic analysis was also used because of the ability to describe the themes and produce a report and other additional advantages (Braun & Clarke, 2006). First, we familiarised ourselves with data, generated initial codes, searched for themes based on SLA, reviewed the themes, defined and named the themes, and finally produced the report (Braun & Clarke, 2006). This was an iterative process which involved so many changes. Data was collected between 30th June and 4th August, 2020. Each interview was transcribed into a Microsoft Word document to simplify the process of analysing the data with the aim of identifying themes.

## 5.   RESULTS AND DISCUSSION

### 5.1 Characteristics of research participants

In terms of demographic characteristics of participants, the findings showed that majority of respondents (56.1% or 14) were aged between 26-35, more respondents' business ownership was proprietorship (92.3% or 23) compared to business partnership (8.0% or 2) The results also revealed that more than half of research participants (68.2% or 17) had basic primary education, followed by 44.3% or 11 of those who were married, while 36.0% or 9 were single. In terms of semi-annual income level, 60.3% or 15 of respondents earned below 100,000 Malawi Kwacha, while few respondents (8.0% or 2) earned more than 200,001 Malawi Kwacha ($1=Mwk850). Further, the results indicated that 32.3% or 8 of respondents were from Uliwa, 40.0% or 10 were from Mlare, and 28.0% or 7 situated at Nyungwe. With respect to number of years microenterprises were in operation, the results indicated that at least half (52.3% or 13) of microenterprises were in operation between 1-5 years.

### 5.3 ICTs and vulnerability context

SLA views people as operating in a context of vulnerability. Information on vulnerabilities can be communicated or accessed using ICTs (Makoza, 2011). In this study, it was revealed that the major sources of vulnerability were unemployment, remoteness, price fluctuations, drought, food insecurity, and migration of clients to urban areas. In addition, operating costs, security and inadequate business skills were also identified as internal setbacks affecting their business activities. *"You know, I sell Maize as my business, and being a subsistence business based in rural areas, the number of people who can buy your products is very small. Again, our area receives sometimes little rainfall, and this make the yield very small. As a result, the prices changes regularly affecting the growth our business" (Resp-17).* The findings from this study were confirmed with prior studies (Qureshi & Xiong, 2017; Duncombe & Heeks, 2005)

Participants acknowledged that ICTs such as mobile phones and social media helped them deal with family emergencies such as death, injury, or other health issues by calling relatives or close friends. ICTs in this case, facilitated women with information and communication about the vulnerabilities in their business activities. "*I use my phone to communicate with my family during emergencies or sometimes when I get sick. "Listening to radios or watching television helps us to understand the weather or the season of the year. This type of information helps us to prepare in advance the type of business activities we need to venture into or forgone" (Resp-12).* However, in this study it was noted that information provided by ICTs were not able to mitigate and prevent all the vulnerabilities that women encountered in their businesses. For instance, it was difficult for ICTs to deal with seasonal price fluctuations, unemployment, and migration of clients to urban areas. Hence, ICTs only provided information to women and helped them to diversify or expand their businesses and sources of income (Osman et al., 2019b; Mbuyisa, 2017; Makoza & Chigona, 2012).

### 5.4 ICTs and utilisation of capital assets
**5.4.1 Human asset:** The study found the majority women did not have business skills before starting their businesses, but rather acquired their skills through informal training through family friends, business workmates, and co-workers. Such prerequisite knowledge and skills developed with time for operating in their businesses. With respect to use of ICTs for business knowledge acquisition, the results found a limited impact. This is what a restraunt owner had to comment: *"Of course, my phone has both internet and WhatsApp facilities, but I do not know how to use them for my business because I do not have skills. I just know how to make calls to my clients and suppliers" (Resp-10).* The findings were consistent with prior study that found no evidence to demonstrate that ICTs were used to acquire business skills for microenterprise in South Africa (Osman, et al., 2019a; Makoza & Chigona, 2012).

**5.4.2 Social asset**: Research participants acknowledged that they belonged to informal social groups such as local village banks, women business clubs, and family business social support. Very few of them belonged to formal organisations such as microloan finances. The problem of not belonging to formal organisations was due to lack of awareness. Other partipants indicated that their businesses were not registered, therefore there was no need to belong to such formal support organisations. *"Those institutions demand that your business must be registered to join them. My business is not registered, there I don't see any importance of joining them" (Resp-10, 6, 1).* Previous studies have echoed similar sentiments that the majority of microenterprises owned by women operate informally and largely belong to informal organisations operating within their localities (Morrison, et al. 2019; Chilimo &Ngulube, 2011). It was further noted that ICTs such mobile phones and social media helped women connect to informal social networks such village banks and women business clubs. ICTs also helped them have access to information on prices of commodities, keep supplier information, and client information. *"Mobile phones and WhatsApp help me to communicate with my business customers and suppliers faster than before. I also use these gadgets to communicate with business friends in our group to share business opportunities" (Resp-7, 25).* This was in line with prior studies that have reported similar findings (Osman et al., 2019b; Lwoga & Chigona, 2020).

**5.4.3 Financial asset:** Makoza and Chigona (2012) note that financial capital for start-up or business expansion remains a big challenge for microenterprise in developing countries. The findings revealed that participants used their savings, self-financing and re-investment of returns to finance their business. Only a few of them indicated that that they had access to informal financial services such as local village banks. Lack of awareness of sources of funding, high interest rates, lack of collateral security were major reasons women could not bother themselves to look for formal financial credit. However, it also emerged that mobile phones helped some women to expand their business through mobile money services. Others indicated that mobile banking helped to save cost such as transport, thereby increasing their income. Similar findings by Chilimo and Ngulube (2011) who found that rural people in Tanzania were using mobile phones to conduct bank transactions, pay suppliers through mobile money and mobile banking.

**5.4.4 Physical asset:** The study found that women needed stable electricity, affordable transport, availability of cheap ICT devices and services such as reduced air time, smartphones, and internet bundles. It was further revealed that apart from physical resources, women stated that they need cheap, timely and relevant information resources required for operating their businesses. Respondents viewed information needs related to business financing, marketing, and promotion and supplier information were needed most. In this regard, participants acknowledged that ICTs such as mobile phones helped them to source supplier information, advertise their products on social media connected to smartphones, store information their customers such as contacts and addresses (Mbuyisa, 2017). It was therefore, evident that ICTs played a limited role on physical assets of women microenterprises. However, what was clear was that there was a need to improve the ICT infrastructure such as establishment of telecentres in rural communities, improve road network and affordable electricity through rural electrification programmes (Qureshi & Xiong, 2017).

**5.4.5 Natural asset:** The study showed that a large number of women did not own natural assets. This is because they were involved in retail business only. Consequently, use of ICTs had little or no impacts to support both physical and natural assets of women microenterprises. Based on the overall impact of ICTs on strengthening the capital assets of microenterprise, the study showed that capital assets are dependent on each other. For instance, increasing the use of ICTs on human capital assets has an impact on financial assets and vice versa. Therefore, increasing use of ICTs on capital assets of women entrepreurship should always be taken as a systemic approach (Mbuyisa, 2017).

**5.5 ICTs and livelihood structures**
The study found that the majority of participants were not aware of such formal support institutions, let alone the policies and regulations that affect their businesses. Only a few of them were aware of policies related to district councils involved in collecting their business operating taxes, SIM registration, and mobile money regulations. These findings were consistent with previous studies (Malanga & Kamanga, 2019; Sife et al., 2017 ).

It also emerged that women interacted with informal organisations such as village banks, women business clubs operating in their localities. The participants indicated that formal support demanded registration of their businesses. Therefore, participants viewed such formal support organisations difficult and not important to their businesses. *"I rarely interact with such organisations because I just think they need a business which is registered and has a lot of operating capital. Yet, my business is very small and this is why I am failing to register it" (Resp-9, 18)*. The findings were consistent with earlier studies (Osman et al., 2019b; Donner & Escobari, 2010). It was also observed that mobile phones and WhatsApp facilitated social and business communication with informal social groups. However, it also emerged that some respondents were reluctant to join informal business social groups due to their culture and jealous from their husbands. *"I have a phone and WhatsApp where I am keeping all contact members in our village bank, do not have problem when we want to meet or share some business information. The only challenge that sometimes I face is that WhatsApp bundle and airtime costs are too high. In addition, as marriage woman, sometimes our husband gives us limits when to use phones"(Resp-14, 17)*.

**5.6 Effects of ICT on livelihood outcomes**
It was revealed that women used mobile banking and mobile payments to diversify their business activities. *"when I found that TNM mobile money is business opportunity, I went to the operator and registered it. Now it is complementing my business. I get monthly income inform of commissions. So, now I have two business complementing each other" (Resp-3,8,11)*. The study found that ICTs increased business earnings through higher income, improved productivity, improved communication, reduction in cost of transport and time. Besides, the study found that ICTs helped women microenterprises towards greater market participation and diversification to high-value business. Similar studies have found that use of ICTs such as mobile phones and internet provide information that help microenterprises to effective use of assets and structures leading to better livelihood outcomes (Osman et al., 2019a; Sife et al., 2017).

# 6. CONCLUSION

The main purpose of the study was to understand the contribution of ICTs to support the livelihoods of microenterprises owned by rural women in Karonga district of northern Malawi. First, the study found that ICTs increased financial capital through branchless banking services of women microenterprises. Mobile banking and mobile money channels available in rural areas provided low cost access and remittance facilities for women microenterprise. This further helped women to diversify and expand their business activities.

Second, Women microenterprises were able to use ICTs such as mobile phones to build their informal business social network groups that strengthened their social, political, and cultural capital assets (Afrah & Fabiha, 2017). ICTs provided women microenterprise with key information and communication channels to support the value chain of their business in rural informal economy. For instance, women microenterprise owners, used mobile phones to access information on prices, buyers and sellers in the local markets and beyond. Furthermore, mobile phones helped women microenterprises to reduce the cost of transactions associated with exchange of information relevant to their business activities. In this regard, dependence on natural or physical capital assets such as roads and transport was reduced. Thus, minimized the travel, administrative and operational costs (Lwoga & Chigona, 2020).

Fourth, the paper found that use of ICTs such as mobile phones enabled women microenterprises increase income revenue and profits; expanding more business opportunities; access to market prices and market information; less dependence on natural/physical capital assets and reduction of risks. Thus, ICTs improved food security at household levels, improved well-being, business growth, productivity, empowerment of women (Noruwana et al., 2018; Mbuyisa, 2017). However, ICT infrastructure, environmental, cultural, personal factors restricted full realisation of livelihood outcomes of the sampled women microenterprises.

Therefore, the findings from this study have both theoretical and policy implications. The use of SLA as an analytical framework, provided empirical evidence on the significant role ICTs play on the livelihoods of women microenterprises in rural setting of a developing country. There is need to improve the ICT infrastructure such as telecentres in rural areas, supported with ICT literacy and business skill programmes tailor-made for women microenterprises. Government should explore alternative sources of power such as solar energy in rural areas. There is need to increase awareness of formal organisations that support SMEs including women microenterprises in the informal sectors. Government should incentivise women microenterprises by offering them free business registration. This will help them gain more access to formal support organisations in the country particularly those that support business management and financing (Lwoga & Chigona, 2020; Sife et al., 2017). The study also has some limitations. Since this study only targeted women microenterprises in one district, it is recommended that similar studies with bigger sample size should be replicated to other parts of the country. Similar studies may employ mixed methods to gain more insights into the study phenomenon. Overall, this study has demonstrated that use of ICTs in microenterprises owned by women vary according to their needs, capacities and opportunities. ICTs can contribute to the business growth, reduction of risks, strengthening capital assets, and improved livelihood outcomes of women microenterprises in different ways. Therefore, developing ICT strategies and tools for enabling women microenterprises to grow must be tailor-based on their business category and the environmental context in which they operate.

## REFERENCES AND CITATIONS